\documentclass[aps,twocolumn,floatfix,groupedaddress,nofootinbib]{revtex4}
\usepackage{csquotes}
\usepackage{graphicx,color}
\usepackage{float}
\usepackage[caption=false]{subfig}
\usepackage{amsmath}
\usepackage{amssymb}
\usepackage{multirow}
\usepackage{dsfont}
\expandafter\let\csname equation*\endcsname\relax
\expandafter\let\csname endequation*\endcsname\relax

\begin{document}

\title{Classical bridge functions in classical and quantum plasma liquids}
\author{F. Lucco Castello$^{1}$, P. Tolias$^{1}$ and T. Dornheim$^{2,3}$}
\affiliation{$^1$ Space and Plasma Physics - Royal Institute of Technology (KTH), SE-10044 Stockholm, Sweden\\
             $^2$ Center for Advanced Systems Understanding (CASUS), D-02826 G\"orlitz, Germany\\
             $^3$ Helmholtz-Zentrum Dresden-Rossendorf (HZDR), D-01328 Dresden, Germany}
\begin{abstract}
\noindent Bridge functions, the missing link in the exact description of strong correlations, are indirectly extracted from specially designed molecular dynamics simulations of classical one-component plasma liquids and accurately parameterized. Their incorporation into an advanced integral equation theory description of Yukawa one-component plasma liquids and a novel dielectric formalism scheme for quantum one-component plasma liquids leads to an unprecedented agreement with available molecular dynamics simulations and new \emph{ab initio} path integral Monte Carlo simulations, respectively.
\end{abstract}
\maketitle

Strongly coupled charged systems, naturally occurring or engineered, are ubiquitous in disparate environments that range from high-energy-density matter\,\cite{introd1,introd2} to soft matter\,\cite{introd3,introdK}. These correlated systems consist of classical point particles or fermions that interact via bare or screened Coulomb pair potentials\,\cite{introd4}. Pivotal to their understanding are three idealized models, whose investigation has led to key physical insights, namely the classical or quantum one-component plasma (OCP)\,\cite{introd5,introd6,DornRev} and the classical Yukawa one-component plasma (YOCP)\,\cite{introd7,UCNPRev}. The OCP and the YOCP liquid states, although squeezed in a rather small portion of the phase diagram between the gas and the crystal states, have proven to be the most theoretically elusive owing to their lack of small parameters that forbid perturbative expansions viable for weak interactions or small vibrations\,\cite{introd8,introd9}. Particular attention has been paid to their structural and thermodynamic properties, since these also constitute input for advanced theoretical descriptions of collective modes\,\cite{introda,introdb,introdN}, dynamical properties\,\cite{introdc} and transport coefficients\,\cite{introdd,introde}. 

In the classical case, the integral equation theory (IET) of liquids constitutes the most accurate alternative to computer simulations for the determination of static pair correlations\,\cite{IETliq1}. For one-component systems, it features two formally exact equations: the Ornstein-Zernike (OZ) integral equation and the non-linear equation\,\cite{IETliq2,IETliq3}
\begin{align}
h(r)&=c(r)+n\int c(r')h(|\boldsymbol{r}-\boldsymbol{r}'|)d^3r'\,,\label{OZequation}\\
g(r)&=\exp\left[-\beta u(r)+h(r)-c(r)+B(r)\right]\,,\label{OZclosure}
\end{align}
with $g(r)$ the radial distribution function (RDF), $h(r)=g(r)-1$ the total correlation function (TCF), $c(r)$ the direct correlation function (DCF), $B(r)$ the bridge function, $u(r)$ the interaction potential, $\beta$ the inverse temperature and $n$ the number density\,\cite{IETliq2}. A $B[h]$ or $B[u]$ functional is required to close the set. In diagrammatic analysis, bridge functions are represented by densely connected irreducible graphs and formally defined by virial-type series that involve Mayer functions or TCFs\,\cite{IETliq4}. Both series converge very slowly and their high-order terms quickly become too complicated to calculate\,\cite{IETliq5}. Moreover, bridge functions lack a probabilistic interpretation and cannot be expressed as ensemble averages of functions that depend on instantaneous particle configurations, implying that they can only be indirectly extracted from simulations; a notoriously difficult task\,\cite{IETliq6,IETliq7,IETliq8,IEMHNC2}. Thus, numerous IET approaches have been developed that approximate the bridge function with varying complexity\,\cite{IETliq9}, the simplest being the hypernetted chain (HNC) approach that drops it altogether, $B(r)\equiv0$,\,\cite{IETliq2}. Indicative of the difficulty of indirect bridge function extraction is the fact that simulation-based bridge function parametrizations are available only for hard spheres (only in the intermediate \& long range)\,\cite{bridgX1}, soft spheres (full range, based on $5$ states)\,\cite{bridgX2} and the OCP (entire range, based on $4$ states and problematic)\,\cite{Iyetomi}. It should be further emphasized that full range bridge function parameterizations are not even available for the paradigmatic liquid of hard spheres and that the well-known analytical hard sphere bridge functions, which have enjoyed wide applications in liquid state theory\,\cite{IETliq1,IETliq2,IETliq3}, are not exact being solutions of the Percus-Yevick approximation\,\cite{bridgX3}.

Here, we extract the classical OCP bridge functions at multiple states, spanning the dense liquid region, from specially designed molecular dynamics (MD) simulations and construct an analytic parametrization. This is incorporated into the recent isomorph-based empirically modified hypernetted chain approach (IEMHNC) based on the excess entropy invariance of YOCP bridge functions\,\cite{IEMHNC1} and into a novel dielectric quantum OCP scheme based on the exact classical-limit correlations. Theoretical predictions are compared with available MD and new path integral Monte Carlo (PIMC) simulations, respectively.

\emph{Classical OCP bridge function extraction.}\,The classical OCP concerns point charges that are immersed in a rigid neutralizing background. The thermodynamic states are fully specified by a single dimensionless quantity, since the non-ideal Helmholtz free energy depends on a specific density-$n$
and temperature-$T$ combination\,\cite{Helmhol}: the coupling parameter $\Gamma=\beta{Q}^2/d$ with $d=[4\pi{n}/3]^{-1/3}$ the Wigner-Seitz radius and $Q$ the particle charge. We focus on moderate densities above the Kirkwood point $\Gamma_{\mathrm{K}}\simeq1.12$\,\cite{Kirkwoo} and prior to the bcc crystallization point $\Gamma_{\mathrm{m}}\simeq171.8$\,\cite{phasedi}. Bridge functions will be indirectly extracted for $17$ state points, $\Gamma=10,20,...170$. The general extraction methodology developed in Ref.\cite{IEMHNC2} can be directly applied to variable softness, purely repulsive or partly attractive, bounded or diverging potentials, but it needs to be modified for the OCP due to the long-range Coulomb interactions. In what follows, we briefly present these peculiarities. Reduced $x=r/d$ units are employed.

Outside the correlation void where $g(x)\simeq0$ ($x>1.2$), bridge functions are indirectly extracted with the \emph{OZ inversion method}\,\cite{IEMHNC2,ocpwitt}. NVT MD simulations are carried out with $N=54872$ particles, $2^{20}$ equilibration time-steps and $2^{23}$ time-steps for statistics leading to $2^{16}$ statistically independent configurations. The long-range interactions are handled with the Ewald sum that is implemented with the particle-particle particle-mesh technique\,\cite{PPPMmur}. The RDF is extracted from histograms with a bin width of $\Delta{x}=0.002$. The Lebowitz-Percus finite-size correction is applied, $g(x)=g_{\mathrm{MD}}(x)(1+\chi_{\mathrm{T}}/N)$\,\cite{leboper}, with $\chi_{\mathrm{T}}$ the reduced isothermal compressibility as calculated from the hypervirial route\,\cite{hypervi}. Fast Fourier Transforms (FFT) are used to compute the static structure factor (SSF) $S(k)$ and Pad\'e approximants are utilized to ensure that the compressibility sum rule is satisfied exactly\,\cite{introd5}. Inverse FFT with long-range decomposition is employed to determine the DCF from the Fourier transformed OZ. Eq.(\ref{OZclosure}) can now be solved for the bridge function.

\begin{figure}
	\centering
	\includegraphics[width=3.20in]{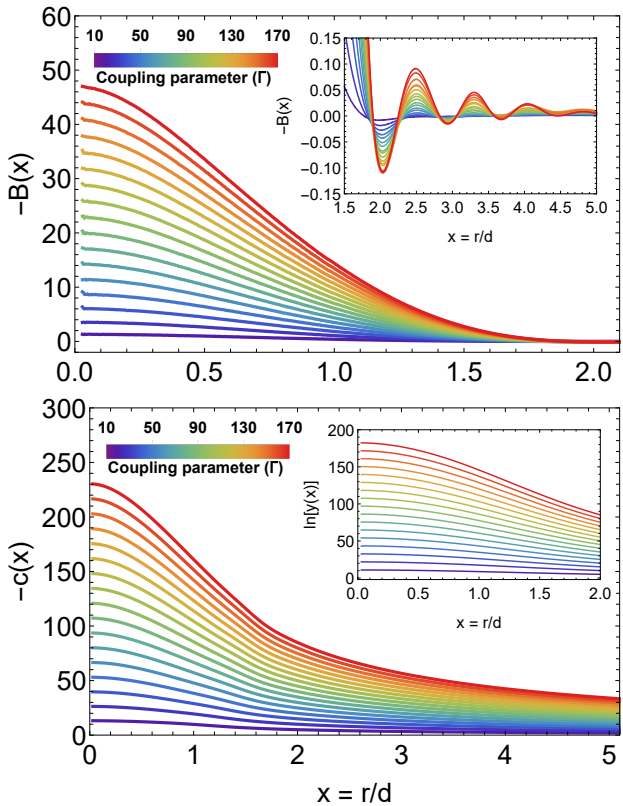}
	\caption{Upper panel: Extracted bridge functions in the monotonic (main) and oscillatory decay range (inset). Lower panel: Extracted direct correlation functions in the non-trivial $B(r)$ range (main) and extracted cavity distribution logarithms in the monotonic range (inset). Results for $17$ OCP state points.}\label{fig:bridgefunctions}
\end{figure}

Within the correlation void ($x<1.4$), bridge functions are indirectly extracted with the \emph{cavity method}\,\cite{IEMHNC2,cavitym}. NVT MD simulations are performed featuring two tagged particles whose artificial pair interactions $\psi(x)=\chi(x)+\phi(x)$ enable sampling of the cavity distribution function (CDF) $y_{\mathrm{sim}}(x)=g^{12}(x)\exp{[\beta\psi(x)]}$. In order to enhance sampling, the correlation void is split into four successive overlapping windows by imposing hard constraints in the tagged pair motion through the $\chi(x)$ component that realizes a smooth potential well. Aiming to achieve uniform sampling, multiple short simulations are run to optimize the $\phi(x)$ component that is determined by supplementing the prescription of Ogata\,\cite{ogatapr} with a linear adder. NVT MD simulations are performed with $N=1000$ particles (useful statistics only from the tagged particles), $2^{20}$ equilibration time-steps and $2^{31}-2^{32}$ time-steps for statistics leading to $2^{24}-2^{25}$ statistically independent configurations. The CDFs of the real and the simulated system are connected by $y(x)=Cy_{\mathrm{sim}}(x)\exp{[(\Gamma/x)\mathrm{erf}(a_{\mathrm{s}}x)]}$ with $a_{\mathrm{s}}$ the Ewald splitting parameter and $C$ determined from CDF continuity. Eq.(\ref{OZclosure}) can now be formulated via $y(r)$ and solved for the bridge function. Our OCP CDFs agree very well with those extracted by Caillol \& Gilles\,\cite{YOCPbri}.

Extraction uncertainties stem exclusively from statistical errors due to the finite simulation duration, since tail errors and implicit size errors are negligible, explicit size errors are corrected and grid errors are minimized\,\cite{IEMHNC2}. All the extracted OCP bridge functions, DCFs and CDFs are featured in Fig.\ref{fig:bridgefunctions} for their entire non-trivial range.

\begin{figure}
	\centering
	\includegraphics[width=3.20in]{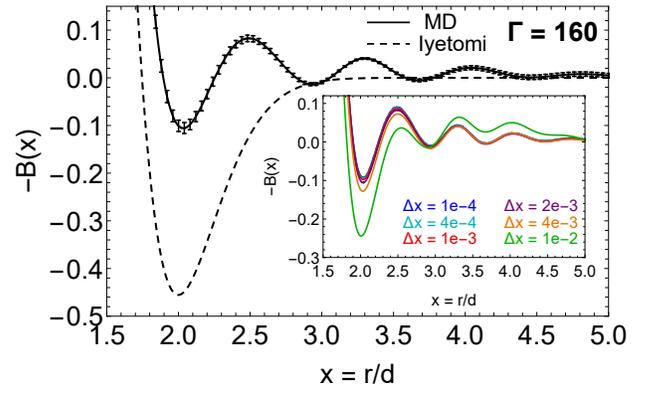}
	\caption{Results for $\Gamma=160$. Our $B(r)$ including uncertainties vs the Iyetomi $B(r)$ within the oscillatory decay range (main). Our $B(r)$ for varying bin widths: large grid errors emerge for $\Delta{x}\gtrsim0.01$ (inset).}\label{fig:griderrors}
\end{figure}

\emph{Classical OCP bridge function parametrization.} To obtain an analytic expression for the entire range, the following procedure was developed. The short range bridge function that exhibits a monotonic behavior is fitted with a fifth order polynomial without a linear term, as suggested by
the short range version of the exact non-linear closure equation $B(x)=\ln{[y(x)]}+c(x)+1$, Widom's general theorem for the CDF $\ln{[y(x)]}=y_0+y_2x^2+y_4x^4$\,\cite{widomth} and the analytic OCP solution of the soft mean spherical approximation for the DCF $c(x)=c_0+c_2x^2+c_3x^3+c_5x^5$ \cite{bipola1,bipola2}. The intermediate range bridge function that exhibits an oscillatory decay is fitted with a combination of exponents and cosines that allow us to exactly capture the first period. These two ranges feature a deliberate overlapping interval, so that the transition region is well described with a sigmoid switching function. Overall,
\begin{align}
&B_{\mathrm{OCP}}(x,\Gamma)=\left[1-f(x)\right]B_{\mathrm{S}}(x,\Gamma)+f(x)B_{\mathrm{I}}(x,\Gamma),\label{OCPbridgefunction}\\
&B_{\mathrm{S}}(x,\Gamma)=s_0(\Gamma)+\textstyle\sum_{i=2}^{5}s_i(\Gamma)x^{i},\nonumber\\
&B_{\mathrm{I}}(x,\Gamma)=l_0(\Gamma)\Gamma^{5/6}\exp{\left[-l_1(\Gamma)(x-1.44)-0.3x^2\right]}\times\nonumber\\&\,\,\,\,\,\,\,\,\,\,\left\{\cos{\left[l_2(\Gamma)(x-1.44)\right]}+l_3(\Gamma)\exp{\left[-3.5(x-1.44)\right]}\right\},\nonumber\\
&f(x)=0.5\left\{1+\mathrm{erf}\left[5.0\left(x-1.5\right)\right]\right\},\nonumber
\end{align}
with\,$s_i(\Gamma)=\textstyle\sum_{j=0}^{3}s_{i}^{j}\Gamma(\ln{\Gamma})^{j},l_i(\Gamma)=\textstyle\sum_{j=0}^{4}l_{i}^{j}\Gamma^{1/6}(\ln{\Gamma})^{j}$ being monotonic functions of $\Gamma$. The $s_{i}^{j},l_{i}^{j}$ coefficients are listed in Table \ref{tab:coefficients}. The functional form of $s_i(\Gamma),l_i(\Gamma)$ has been inspired from exact low-$\Gamma$ expansions of the excess OCP internal energy beyond the Debye-H\"uckel term\,\cite{expans1,expans2}. For all states, the fit is near-exact within $0\leq{x}\leq3$ but it fails to describe higher order damped oscillations that arise up to $x\simeq5$ near the melting point.

\begin{table}
	\caption{Fit parameters of the $B(r)$ parametrization, Eq.(\ref{OCPbridgefunction}).}\label{tab:coefficients}
	\centering
	\begin{tabular}{cccccc}
		\hline
		        & $j=0$     & $j=1$     & $j=2$     & $j=3$       & $j=4$      \\ \hline
		$s_0^j$ & 0.076912  & -0.10465  & 0.0056629 & 0.00025656  &    N/A     \\
		$s_2^j$ & 0.068045  & -0.036952 & 0.048818  & -0.0048985  &    N/A     \\
		$s_3^j$ & -0.30231  & 0.30457   & -0.11424  & 0.0095993   &    N/A     \\
		$s_4^j$ & 0.25111   & -0.26800  & 0.082268  & -0.0064960  &    N/A     \\
		$s_5^j$ & -0.061894 & 0.066811  & -0.019140 & 0.0014743   &    N/A     \\
		$l_0^j$ & 0.25264   & -0.31615  & 0.13135   & -0.023044   & 0.0014666  \\
		$l_1^j$ & -12.665   & 20.802    & -9.6296   & 1.7889      & -0.11810   \\
		$l_2^j$ & 15.285    & -14.076   & 5.7558    & -1.0188     & 0.06551    \\
		$l_3^j$ & 35.330    & -40.727   & 16.690    & -2.8905     & 0.18243    \\  \hline
	\end{tabular}
\end{table}

OCP bridge functions were earlier extracted and parameterized by Iyetomi and coworkers\,\cite{Iyetomi}. Their bridge function has a number of deficiencies:\,(\textbf{a}) The short range was determined with extrapolations based on the Widom theorem\,\cite{widomth} and the exact Jancovici order $x^2$ result\,\cite{Jancovi}. (\textbf{b}) The extraction concerned only four state points, \emph{i.e.} $\Gamma=10,40,80,160$. (\textbf{c}) The parametrization consisted of a high-order polynomial multiplied by an exponentially decaying function, which lead to a single extremum curve without oscillatory decay. (\textbf{d}) The RDF histograms had a relatively large bin width of $\Delta{x}=0.04$, which translates to large grid errors mainly near the $B(r)$ extremum. Although their short range extrapolation method turned out to be accurate, the other deficiencies are important. In particular, their large grid errors are revealed in Fig.\ref{fig:griderrors}.

\emph{Application to classical plasma liquids.} The YOCP concerns point charges that interact via the Yukawa potential $u(r)=(Q^2/r)\exp{(-r/\lambda_{\mathrm{s}})}$ being embedded in a polarizable neutralizing background, with $\lambda_{\mathrm{s}}$ a shielding length. The thermodynamic states are specified by two dimensionless quantities\,\cite{CompRev}: the coupling and screening parameters, $\Gamma$ and $\kappa=d/\lambda_{\mathrm{s}}$. The OCP is recovered as $\kappa\to0$. We focus on moderate densities above the Kirkwood line\,\cite{Kirkwoo} and prior to the bcc/fcc crystallization\,\cite{phasedi}.

\begin{figure}
	\centering
	\includegraphics[width=2.85in]{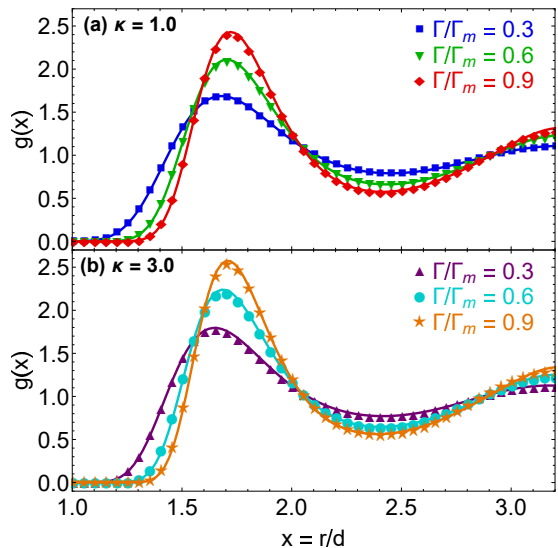}
	\caption{Radial distribution functions resulting from MD simulations (symbols) and the updated IEMHNC approach (solid lines): (a) $\kappa=1$ and $\Gamma/\Gamma_{\mathrm{m}}=0.3$ (blue), 0.6 (green), 0.9 (red) with $\Gamma_{\mathrm{m}}=220.18$, (b) $\kappa=3$ and $\Gamma/\Gamma_{\mathrm{m}}=0.3$ (purple), 0.6 (cyan), 0.9 (orange) with $\Gamma_{\mathrm{m}}=1234.51$.}\label{fig:rdfs}
\end{figure}

It has recently been demonstrated that the YOCP exhibits very strong correlations between its virial and potential energy constant volume thermal equilibrium fluctuations in the entire dense liquid region of its phase diagram\,\cite{isomor1}. Thus, the YOCP belongs to the class of R-simple systems and possesses isomorphic lines, \emph{i.e.} phase diagram curves of constant excess entropy along which a set of structural and dynamic properties are nearly invariant when expressed in properly reduced units\,\cite{isomor2,isomor3}. In particular, the isomorphic lines are nearly parallel to the melting line\,\cite{isomor4} and accurately parameterized by\,\cite{isomor1,Khrapak}
\begin{equation}
\Gamma_{\mathrm{ISO}}(\Gamma,\kappa)=\Gamma{e}^{-\alpha\kappa}\left[1+\alpha\kappa+(1/2)(\alpha\kappa)^2\right]=\mathrm{const.}\label{YOCPisomorph}
\end{equation}
with $\alpha=(4\pi/3)^{1/3}$. Our recent YOCP simulations\,\cite{IEMHNC2} proved that the reduced-unit bridge functions of R-simple systems are isomorph invariant, validating the conjecture of Ref.\cite{IEMHNC1}. Thus, given the Eq.(\ref{YOCPisomorph}) mapping of configurational adiabats, the OCP bridge functions constitute the basis for the construction of YOCP bridge functions via
\begin{equation}
B_{\mathrm{YOCP}}(x,\Gamma,\kappa)=B_{\mathrm{OCP}}[x,\Gamma_{\mathrm{ISO}}(\Gamma,\kappa)]\,.\label{YOCPbridgefunction}
\end{equation}
This IET closure amounts to the IEMHNC approach that was earlier combined with the Iyetomi OCP bridge function and applied to the YOCP\,\cite{IEMHNC1}. Detailed benchmarking activities against simulation results revealed that this early IEMHNC version could reproduce the YOCP thermodynamic properties within $0.5\%$ and the YOCP structural properties within $1.5\%$ inside the first coordination cell\,\cite{IEMHNC3}; an excellent performance comparable to that of the variational modified hypernetted chain approach\,\cite{IEMHNC4} (VMHNC) that is $10-100\times$ more computationally costly.

\begin{figure*}[t]
	\centering
	\includegraphics[width=5.45in]{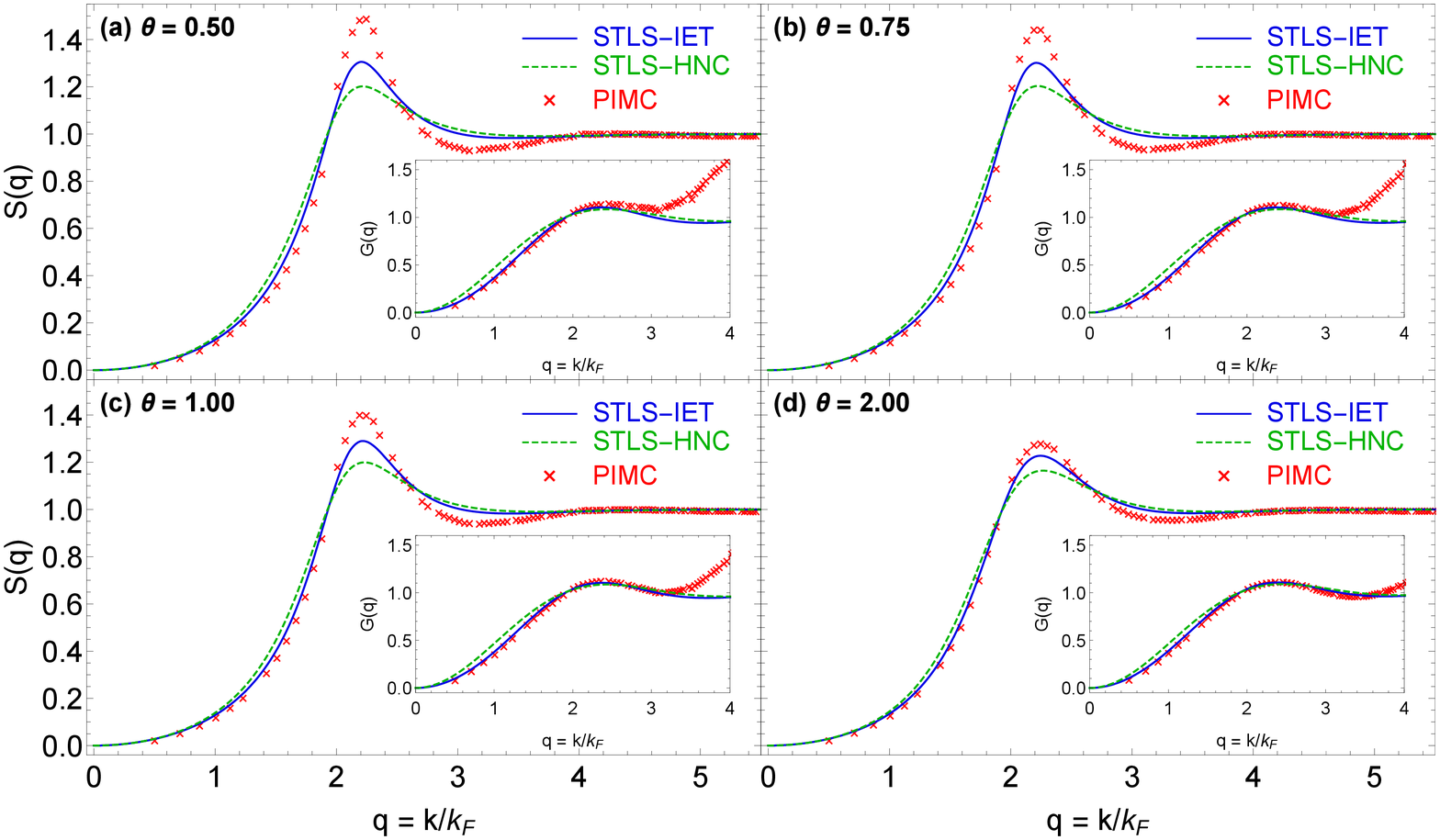}
	\caption{Static structure factors (main) and static local field corrections (inset) resulting from the PIMC simulations (red crosses), HNC-based scheme (dashed green lines) and IET-based scheme (solid blue lines). Results for $r_{\mathrm{s}}=100$ and $\theta=0.5,0.75,1,2$.}\label{fig:quantum}
\end{figure*}

\begin{figure}[t]
	\centering
	\includegraphics[width=3.45in]{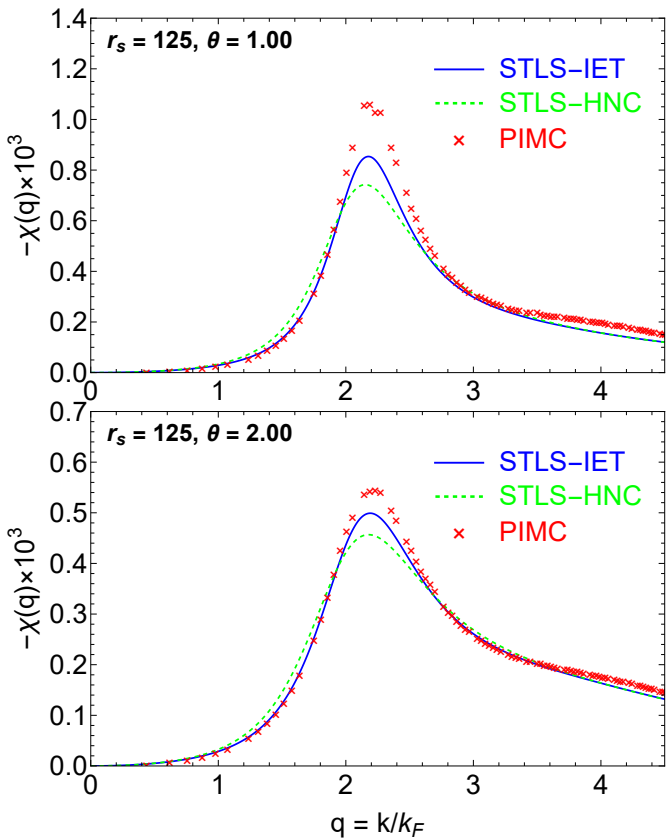}
	\caption{Static density-density responses $\chi(\boldsymbol{k})=\chi(\boldsymbol{k},0)$ resulting from the PIMC simulations (red crosses), HNC-based scheme (dashed green lines) and IET-based scheme (solid blue lines). Results for $r_{\mathrm{s}}=125$ and $\theta=1,2$.}\label{fig:staticdensityresponse}
\end{figure}

The updated IEMHNC approach is obtained by combining Eqs.(\ref{OZequation},\ref{OZclosure}) with Eqs.(\ref{OCPbridgefunction},\ref{YOCPisomorph},\ref{YOCPbridgefunction}). This set is solved with Picard iterations in Fourier space combined with mixing and long-range decomposition techniques (when $\kappa<1$). Comparison with extended simulations\,\cite{phasedi,YOCPth1,YOCPth2} reveals that the updated IEMHNC version reproduces YOCP thermodynamic and structural properties within $0.5\%$ in the whole dense liquid region; an unprecedented accuracy on par with that of modern simulations. This is highlighted in the graphical comparison between IEMHNC- \& MD-generated RDFs featured in Fig.\ref{fig:rdfs}. The superiority of our updated IEMHNC approach over the VMHNC approach\,\cite{IEMHNC4,IEMHNC5} and another advanced IET approach\,\cite{IEMHNC6} is confirmed in the Supplemental Material\,\cite{supplem}.

\emph{Application to quantum plasma liquids.} The quantum OCP concerns electrons immersed in a rigid ionic neutralizing background. In the unpolarized case of equal spin-up \& -down electrons, thermodynamic states are specified by two dimensionless quantities\,\cite{BoniRev}: the Brueckner parameter $r_{\mathrm{s}}=d/a_{\mathrm{B}}$ with $a_{\mathrm{B}}=\hbar^2/(m_{\mathrm{e}}e^2)$ the first Bohr radius and the degeneracy parameter $\theta=k_{\mathrm{B}}T/E_{\mathrm{F}}$ with $E_{\mathrm{F}}=[(3\pi^2n)^{2/3}/2](\hbar^2/m_{\mathrm{e}})$ the Fermi energy w.r.t spin-up electrons ($\hbar$ is the reduced Planck constant,\,$e$ the elementary charge). We focus on high degeneracy ($\theta\sim1$) moderate density ($r_{\mathrm{s}}\gtrsim20$) post warm dense matter\,\cite{WDMbook} but prior to Wigner crystallization\,\cite{Wigner1,Wigner2}, where correlations are strong but quantum effects remain important.

In linear response theory, the exact density-density response function $\chi(\boldsymbol{k},\omega)$ can always be expressed in terms of the ideal (Lindhard) density response $\chi_0(\boldsymbol{k},\omega)$ and the unknown dynamic local field correction $G(\boldsymbol{k},\omega)$ (LFC)
\begin{equation}
\chi(\boldsymbol{k},\omega)=\frac{\chi_0(\boldsymbol{k},\omega)}{1-U(\boldsymbol{k})\left[1-G(\boldsymbol{k},\omega)\right]\chi_0(\boldsymbol{k},\omega)}\,,\label{densityresponseDLFC}
\end{equation}
with $U(\boldsymbol{k})=4\pi{e}^2/k^2$ the regularized Fourier transform of the Coulomb pair potential\,\cite{Ichibok}. In addition, frequency integration of the quantum fluctuation-dissipation theorem (FDT) and analytic continuation of $\chi(\boldsymbol{k},\omega)$ to the complex plane lead to a static structure factor $S(\boldsymbol{k})$ (SSF) relation that involves the Matsubara summation
\begin{equation}
S(\boldsymbol{k})=-\frac{1}{{n}\beta}\displaystyle\sum_{l=-\infty}^{\infty}\widetilde{\chi}(\boldsymbol{k},\imath\omega_l)\,,\label{Matsubaraseries}
\end{equation}
with $\widetilde{\chi}(\boldsymbol{k},z)$ the complex-valued density-density response function, $\omega_l=2\pi{l}/(\beta\hbar)$ the Matsubara frequencies\,\cite{IchiMat}. Dielectric schemes approximate the LFC as a SSF functional $G\equiv{G}[S]$, leading to self-consistent approaches\,\cite{Ichibok,DornRev}. Rigorous schemes that include quantum effects on the random phase approximation level and treat correlations classically (such as the Singwi-Tosi-Land-Sj\"olander [STLS] scheme\,\cite{STLSgro,STLSfin}) and semi-empirical schemes that employ asymptotic limits and embody simulation results (see the effective static approximation\,\cite{ESAlett,ESApape}) are based on a frequency-averaged simplification, $G(\boldsymbol{k},\omega)\equiv{G}(\boldsymbol{k})$.

A recently proposed scheme is singled out that belongs to the first group and treats strong correlations within the classical HNC approach\,\cite{HNCSTLS,HNCPIMC}. This HNC-based scheme combines the classical FDT, OZ equation and HNC non-linear equation to generate a frequency averaged $G[S]$ functional. Systematic comparison with PIMC results for moderate\,\cite{HNCSTLS} and strong coupling\,\cite{HNCPIMC} has revealed that the HNC-based scheme is superior to other dielectric schemes. To be specific, when $r_{\mathrm{s}}\in[20,100]\cap\theta\in[0.5,4]$, its interaction energy predictions are accurate within $1.2\%$ due to favorable error cancellations in the SSF integration\,\cite{HNCPIMC} while its structural predictions are quite accurate for the SSF/LFC peak positions but significantly underestimate the SSF / LFC peak heights\,\cite{HNCPIMC}. Considering that such a deficiency is also characteristic of the fully classical HNC approach\,\cite{HNCcla1,HNCcla2}, it is expected that incorporation of the bridge function will lead to significant improvements. To this end, we generalize the HNC-based scheme to a novel IET-based scheme including our classical bridge function. The $G[S]$ functional reads as
\begin{align}
G&(\boldsymbol{k})=\frac{{B}(\boldsymbol{k})}{\beta{U}(\boldsymbol{k})}-\frac{1}{n}\int\,\frac{d^3q}{(2\pi)^3}\frac{\boldsymbol{k}\cdot\boldsymbol{q}}{q^2}\left[S(|\boldsymbol{k}-\boldsymbol{q}|)-1\right]\nonumber\\&\times\left\{-\frac{B(\boldsymbol{q})}{\beta{U}(\boldsymbol{q})}+1-\left[G(\boldsymbol{q})-1\right]\left[S(\boldsymbol{q})-1\right]\right\}.\label{IETSTLSLFC}
\end{align}
Use of the classical OCP bridge function necessitates the mapping of the quantum states ($r_{\mathrm{s}},\theta$) to classical states ($\Gamma$) via $\Gamma=2\lambda^2(r_{\mathrm{s}}/\theta)$ with $\lambda^3=4/(9\pi)$. The Eqs.(\ref{densityresponseDLFC},\ref{Matsubaraseries},\ref{IETSTLSLFC}) form a closed set that is solved numerically. The computational cost drastically decreases by converting the triple to a double integral in Eq.(\ref{IETSTLSLFC}) with two-center bipolar coordinates\,\cite{bipola1}. Faster Matsubara summation convergence is achieved by isolating the Hartree-Fock SSF in Eqs.(\ref{densityresponseDLFC},\ref{Matsubaraseries}) and faster high-$k$ convergence for the double integral is achieved by isolating the STLS LFC in Eq.(\ref{IETSTLSLFC}).

To validate the IET-based scheme, new PIMC simulations have been performed with $N=100$ electrons for 16 states ($50\leq{r}_{\mathrm{s}}\leq200$, $0.5\leq\theta\leq2$). For these states, the fermion sign problem is weak owing to the prevalence of strong correlations and the standard PIMC method suffices to obtain accurate results\,\cite{DornRev,FSPtobi}. Interaction energy finite-size errors, that stem from the omission of the long-wavelength contribution in the discretized integral, are corrected applying the perfect screening sum rule\,\cite{DornRev,HNCPIMC}.

\begin{table}[t]
	\caption{Interaction energy $\widetilde{u}=(\pi\lambda{r_{\mathrm{s}}})^{-1}\int_0^{\infty}\left[S(x)-1\right]dx$ (in Hartree units) of the unpolarized electron liquid: PIMC, HNC-based and IET-based results. All PIMC simulations are new except from the first $6$ states\,\cite{HNCPIMC}.}\label{tab:internalenergies}
	\centering
	\begin{tabular}{ccccccc}
	\hline
	$r_{\mathrm{s}}$ & $\theta$ & $\widetilde{u}$ &  $\widetilde{u}$ &  $e_{\mathrm{HNC}}$ &  $\widetilde{u}$   &   $e_{\mathrm{IET}}$    \\
	                 &          & PIMC            &  HNC-based       &  ($\%$)             &  IET-based         &   ($\%$)                \\ \hline
     100             &   0.50   & -0.00825500     & -0.00815866      &     1.167           & -0.00822181        &        0.402            \\
     100             &   0.75   & -0.00824570     & -0.00816490      &     0.980           & -0.00822544        &        0.246            \\
     100             &   1.00   & -0.00823490     & -0.00816618      &     0.834           & -0.00822559        &        0.113            \\
     100             &   2.00   & -0.00817650     & -0.00812905      &     0.580           & -0.00819066        &        0.173            \\
     100             &   4.00   & -0.00800623     & -0.00796833      &     0.473           & -0.00803143        &        0.315            \\
      50             &   0.50   & -0.01600700     & -0.01589841      &     0.678           & -0.01603510        &        0.176            \\
      60             &   0.50   & -0.01345310     & -0.01334804      &     0.781           & -0.01346014        &        0.052            \\
      70             &   0.50   & -0.01161175     & -0.01150938      &     0.882           & -0.01160390        &        0.068            \\
      80             &   0.50   & -0.01021937     & -0.01012012      &     0.971           & -0.01020149        &        0.175            \\
      90             &   0.50   & -0.00912862     & -0.00903293      &     1.048           & -0.00910415        &        0.268            \\
     110             &   0.50   & -0.00752642     & -0.00744012      &     1.147           & -0.00749675        &        0.394            \\
     125             &   0.50   & -0.00665421     & -0.00657377      &     1.209           & -0.00662268        &        0.474            \\
     125             &   0.75   & -0.00665053     & -0.00657838      &     1.085           & -0.00662556        &        0.442            \\
     125             &   1.00   & -0.00664336     & -0.00657999      &     0.954           & -0.00662647        &        0.254            \\
     125             &   1.50   & -0.00662535     & -0.00657432      &     0.770           & -0.00662112        &        0.064            \\
     125             &   2.00   & -0.00660298     & -0.00655900      &     0.666           & -0.00660712        &        0.063            \\
     150             &   0.50   & -0.00558177     & -0.00550821      &     1.318           & -0.00554797        &        0.606            \\
     150             &   1.00   & -0.00557134     & -0.00551337      &     1.040           & -0.00555132        &        0.359            \\
     200             &   0.50   & -0.00422244     & -0.00416445      &     1.373           & -0.00419373        &        0.680            \\
     200             &   1.00   & -0.00421710     & -0.00416813      &     1.161           & -0.00419559        &        0.510            \\  \hline
	\end{tabular}
\end{table}

In terms of static structure, a comparison reveals that: (\textbf{a}) The IET-based scheme substantially improves the SSF peak magnitude and also marginally improves the SSF peak position predictions of the HNC-based scheme. (\textbf{b}) IET-based predictions for the LFC are remarkably accurate especially for $k/k_{\mathrm{F}}\leq2$ with $k_{\mathrm{F}}=(3\pi^2n)^{1/3}$ the Fermi wavevector. Notice that both schemes converge towards the constant value $G(k\to\infty)=1-g(0)\simeq1$, with $g(0)$ the radial distribution function at contact\,\cite{IchiMat}. This is a direct consequence of the $G(\boldsymbol{k},\omega)\equiv{G}(\boldsymbol{k})$ assumption that essentially introduces a frequency-averaged LFC. At the ground state $\theta=0$, the exact static LFC $G(\boldsymbol{k},\omega\to0)$ has been proven to be parabolically divergent with the $k^2$ pre-factor determined by the exchange-correlation contribution to the kinetic energy\,\cite{asyWDM1}. At finite temperatures $\theta>0$, although a rigorous proof is lacking, PIMC simulations have revealed that this divergence persists\,\cite{HNCPIMC,asyWDM2}. Given the above, dielectric schemes that are based on the $G(\boldsymbol{k},\omega)\equiv{G}(\boldsymbol{k})$ assumption, even empirical schemes that incorporate exact PIMC data\,\cite{ESAlett,ESApape}, are bound to exhibit increasingly large LFC deviations for $k/k_{\mathrm{F}}\gtrsim3$. (\textbf{c}) The IET-based scheme drastically improves the $\chi(\boldsymbol{k},0)$ static density-density response predictions of the HNC-based scheme. Improvements include the extremum magnitude, while the predictions for $k/k_{\mathrm{F}}\leq2$ are nearly exact. The conclusions are valid for all simulated states, see Figs.\ref{fig:quantum},\ref{fig:staticdensityresponse} for examples. In terms of thermodynamics, the favorable error cancellation persists, thus, the IET-based interaction energies are accurate within $0.7\%$ compared to $1.4\%$ for the HNC-based scheme, see Table \ref{tab:internalenergies}. It is also noted that the IET- and HNC-based interaction energies are much more accurate than those of the classical mapping method\,\cite{PDWmap1,PDWmap2}, see the Supplemental Material\,\cite{supplem}.

\emph{Discussion.} We performed specially designed MD simulations to indirectly extract the bridge functions of classical OCP liquids. Systematic extractions led to an accurate parametrization that was embedded in the recently proposed IEMHNC integral equation theory approach for classical YOCP liquids and a novel IET-based dielectric scheme for quantum OCP liquids. Extensive PIMC simulations were carried out to facilitate benchmarking. For both liquids, the structural and thermodynamic properties were predicted with unprecedented precision.

Classical OCP bridge functions can be used to explore the limits of other existing theoretical approaches. For the YOCP, the classical OCP liquid can constitute the reference system of the VMHNC approach in place of the Percus-Yevick hard-sphere liquid with the effective coupling parameter determined by minimizing an approximate free energy functional\,\cite{VMHNCRo}. For the quantum OCP, classical OCP bridge functions can be used in classical mapping approaches in place of hard-sphere bridge functions\,\cite{mapping}. Moreover, although this Letter is dedicated to 3D one-component plasma liquids, the bridge function indirect extraction technique, IEMHNC integral equation theory approach and IET-based dielectric scheme can be directly extended to 2D and multi-component systems.

Finally, it should be pointed out that knowledge of the OCP bridge function is directly transferable to any model system (hard sphere, Lennard-Jones, inverse power law, Gaussian core). To be more specific, a substantial part of the success of integral equation theory and fundamental measure classical density functional theory is owed to the bridge function universality ansatz of Rosenfeld-Ashcroft \,\cite{outro01} \& bridge functional universality ansatz of Rosenfeld \cite{outro02,outro03}. These celebrated conjectures state that bridge functions and functionals have reduced-unit forms that are weakly dependent on the details of the pair interactions. In other words, the exact knowledge of the bridge function or functional of one model liquid, supplemented with a powerful variational principle that determines an optimal state correspondence between systems, can lead to the accurate yet approximate knowledge of the bridge function or functional of any other model liquid.

\emph{Acknowledgments.} This work was partly funded by the Swedish National Space Agency under grant no.\,143/16. This work was also partly funded by the Center of Advanced Systems Understanding (CASUS) that is financed by Germany's Federal Ministry of Education and Research (BMBF) and the Saxon Ministry for Science, Culture and Tourism (SMWK) with tax funds on the basis of the budget approved by the Saxon State Parliament. The MD simulations were carried out on resources provided by the Swedish National Infrastructure for Computing (SNIC) at the NSC (Link{\"o}ping University) that is partially funded by the Swedish Research Council under grant agreement no.\,2018-05973. The PIMC simulations were carried out at the Norddeutscher Verbund f\"ur Hoch- und H\"ochstleistungsrechnen (HLRN) under grant no.\,shp00026 and on a Bull Cluster at the Center for Information Services and High Performance Computing (ZIH) at Technische Universit\"at Dresden.

\end{document}